# THE IMPLICATIONS OF A COSMOLOGICAL INFORMATION BOUND FOR COMPLEXITY, QUANTUM INFORMATION AND THE NATURE OF PHYSICAL LAW

P.C.W. Davies

BEYOND: Center for Fundamental Concepts in Science, Arizona State University
P.O. Box 876505, Tempe, AZ 85287-6505

*Whereof one cannot speak, thereof one must remain silent.*

Ludwig Wittgenstein[1]

*Is our universe a polynomial or an exponential place?*

Scott Aaronson[2]

**Abstract**

The finite age of the universe and the existence of cosmological horizons provides a strong argument that the observable universe represents a finite causal region with finite material and informational resources. A similar conclusion follows from the holographic principle. In this paper I address the question of whether the cosmological information bound has implications for fundamental physics. Orthodox physics is based on Platonism: the laws are treated as infinitely precise, perfect, immutable mathematical relationships that transcend the physical universe and remain totally unchanged by physical processes, however extreme. If instead the laws of physics are regarded as akin to computer software, with the physical universe as the corresponding hardware, then the finite computational capacity of the universe imposes a fundamental limit on the precision of the laws and the specifiability of physical states. That limit depends on the age of the universe. I examine how the imprecision of the laws impacts on the evolution of highly entangled states and on the problem of dark energy.

# 1. What are the laws of physics?

Gregory Chaitin is undoubtedly one of the most profound thinkers of our time. I have drawn on his work at several stages in my own career development, and especially in formulating the ideas that follow. It is an honor to contribute to this volume to celebrate Chaitin's important insights into mathematics, computing and physical reality.

I should like to start with a quotation from Chaitin's recent book *Meta-Math*: "Why should I believe in a real number if I can't calculate it, if I can't prove what its bits are, and if I can't even refer to it? …The real line from 0 to 1 looks more and more like a Swiss cheese."[3] In other words, the real line is a useful fiction, an unattainable idealization. The question I wish to address here is how this sweeping conclusion impacts on my own field of theoretical physics and cosmology. The real line, its extension to the complex plane, together with the related properties of differentiability, play an absolutely central role in theoretical physics, on account of the fact that all the known fundamental laws of physics are expressed in terms of differentiable functions defined over the set of real or complex numbers. So I want to start by asking a very basic, but surprisingly little addressed question: What are the laws of physics and where do they come from? The subsidiary question, Why do they have the form that they do? I have discussed in detail elsewhere.[4]

First let me articulate the orthodox position, adopted by most theoretical physicists, which is that the laws of physics are immutable: absolute, eternal, perfect mathematical relationships, infinitely precise in form. The laws were imprinted on the universe at the moment of creation, i.e. at the big bang, and have since remained fixed in both space and time. The properties of the physical universe depend in an obvious way on the laws of physics, but the basic laws themselves depend not one iota on what happens in the physical universe. There is thus a fundamental asymmetry: the states of the world are affected by the laws, but the laws are completely unaffected by the states – a dualism that goes back to the foundation of physics with Galileo and Newton. The ultimate source of the laws is left vague, but it is tacitly assumed to transcend the universe itself, i.e. to lie beyond the physical world, and therefore beyond the scope of scientific inquiry. The proper task of the physicist, it is often said, is to discover the forms of the laws using reason and experiment, adopt them pragmatically, and get on with the job of determining their consequences. Inquiry into their origin is discouraged as a quasi-religious quest.

The orthodox view of the nature of physical laws conforms well to the mathematical doctrine of Platonism. Plato regarded mathematical forms and relationships as enjoying a real existence in an otherworldly realm, where mathematicians come upon them in a voyage of intellectual discovery. A Platonist regards mathematics as possessing an existence independent of the physical universe, rather than being a product of the human brain. An essential quality of the Platonic heaven is that the mathematical forms it contains are perfect. For example, circles are exactly round, in contrast to circles in the physical universe, which are always flawed approximations to the idealized Platonic forms.

Most theoretical physicists are by temperament Platonists. They envisage the laws of physics too as perfect idealized mathematical relationships and operations that really exist, located in an abstract realm transcending the physical universe. I shall call this viewpoint physical Platonism to distinguish it from mathematical Platonism. Newton was a physical Platonist, and cast his laws of mechanics and gravitation in terms of what we would now call real numbers and differentiable functions. Taking Newton's laws seriously implies accepting infinite and infinitesimal quantities, and arbitrary precision. The idealized, Platonic notion of the laws of physics reached its zenith with the famous claim of Laplace, concerning an omniscient demon. Laplace pointed out that the states of a closed deterministic system, such as a finite collection of particles subject to the laws of Newtonian mechanics, are completely fixed once the initial conditions are specified[5]

"We may regard the present state of the universe as the effect of its past and the cause of its future. An intellect which at any given moment knew all of the forces that animate nature and the mutual positions of the beings that compose it, if this intellect were vast enough to submit the data to analysis, could condense into a single formula the movement of the greatest bodies of the universe and that of the lightest atom; for such an intellect nothing could be uncertain and the future just like the past would be present before its eyes."

If Laplace's argument is taken seriously, on the assumptions adopted, then everything that happens in the universe, including Laplace's decision to write the above words, my decision to write this article, Chaitin's beautiful work on Omega, etc. are all preordained. The information about these events is already contained in the state of the universe at any previous time. To get some idea of the demon's gargantuan task, note the following. If the demon overlooked the gravitational force of a single electron located at the edge of the observable universe, then his prediction for the motion of a given molecule of air in your living room would be rendered completely uncertain after only 12 intermolecular collisions.[6] This arresting example reveals how exquisitely sensitive to error predicting the future can be.

Laplace's vignette is based on classical mechanics, and is usually dismissed by invoking quantum mechanics, or arguing that the universe is an open system, but this misses the point. The real absurdity in Laplace's statement is its implicit reliance on physical Platonism extrapolated to a staggering degree, made without any experimental foundation whatever.

In spite of the fact that we now know Newtonian mechanics is only an approximation, physical Platonism remains the dominant philosophy among theoretical physicists. The project of quantum cosmology, for example, is predicated on the assumption that the laws of quantum mechanics and general relativity exist independently of the universe, and may therefore be invoked to explain how the universe came to exist from nothing. In the fashionable subject of string/M theory, the string Lagrangian, or whatever else serves to determine the unified dynamics, is assumed to somehow "already exist", so that from it may (one day) flow an explanation for space, time, matter and force.

**2. Laws as software**

A completely different view of the relationship between mathematics and physics comes from Chaitin's development of algorithmic information theory, from which he was drawn to the conclusion, "A scientific theory is like a computer program that predicts our observations."[7] For example, in Newtonian mechanics the initial positions and momenta of a system of particles serve as input data, the laws of mechanics are the program, and the final state of the particles at some later time of interest corresponds to the output. In this manner, the universe processes information automatically as it evolves. So we might envisage the laws of physics in terms of software, as a grand cosmic computer program. This shift of perspective, simple though it may be, has profound implications, which are immediately apparent when we ask what is the hardware on which the cosmic software is being run? The answer is, of course, the universe itself. And by this I mean the real, physical universe. I am not referring to some imaginary cosmic hardware in a Platonic heaven, but the real universe we observe. The significance of this last point is that the real universe might very well be *finite*, that is, have finite resources and age, and thus be subject to restrictions on what it can accomplish in regards to computation.

Why might the universe be finite in resources? What matters for computational purposes is not the spatial extent of the universe, but the number of physical degrees of freedom located in a causally connected region. Information processed in causally disconnected parts of space cannot be considered as belonging to the same "program." In the standard cosmological models, the region of the universe to which we have causal access at this time is limited by the finite speed of light and finite age of the universe (since the big bang). That is, there exists a "particle horizon" in space, measuring some billions of light years across at this time. The region within our particle horizon contains about $10^{80}$ particles of matter, and about $10^{90}$ photons and neutrinos. If the system is treated quantum mechanically, with information encoded in discrete bits (e.g. spin up, spin down), then the maximum number of bits of information contained in a horizon volume at this time is about $10^{122}$ according to Seth Lloyd.[8] His calculation takes into account the gravitational degrees of freedom too. If the universe is uniform, any other causal region would possess a similar upper bound. Thus we may write

$$I_{universe} \leq 10^{122}. \qquad (1)$$

The bound (1) is not fixed, but grows with time as the horizon expands and encompasses more particles:

$$I_{universe} \propto t^2. \qquad (2)$$

It is a simple matter, using quantum mechanics and thermodynamics, to also calculate the maximum amount of information that could have been processed (i.e. the total number of possible bit flips) in our causal region since the origin of the universe. The answer comes out again of order $10^{122}$, taking into account Eq. (2), i.e. the fact that the causal region was smaller in the past and so encompassed less particles.

A similar information bound may be derived from an entirely different line of argument, exploiting the link between physics and information discovered by Bekenstein[9] and Hawking[10] when applying quantum mechanics to black holes. They found that an uncharged, non-rotating black hole possesses entropy *S* given by

$$S = 4\pi kGM^2/\hbar c^3 = \tfrac{1}{4}A, \qquad (3)$$

where *M* and *A* are the mass and area of the black hole respectively, and the other symbols have their usual meanings as various fundamental constants of nature.

The fact that the entropy is a function of black hole *area*, as opposed to volume, is deeply significant. In the case of a laboratory gas, for example, entropy is additive: twice the volume of a (homogeneous) gas will have twice the entropy. Evidently, when gravitation enters the picture, the rules of the game change fundamentally. Entropy can been regarded as a measure of information $I$ (or information loss), through the relationship

$$S = k\log_2 I \qquad (4)$$

so the Bekenstein-Hawking formula (3) relates the total information content of a region of space to the area of the surface encompassing that volume. The information inside a black hole is lost because an observer in the external region cannot access it on account of the fact that the surface of the hole is an event horizon. (There remains an unresolved issue about whether the information is permanently lost, or just rendered inaccessible until the black hole eventually evaporates. I shall not consider that topic further in this chapter.) A useful way to think about Eq. (3) is to define the Planck length $L_P \equiv (G/\hbar c^3)^{\frac{1}{2}}$ as a fundamental unit, and note that, using Eq. (4), the information of the black hole is simply one quarter of the horizon area in Planck units.

Early on, Bekenstein sought to generalize his result by postulating that Eq. (1) serves as a *universal* bound on entropy (or information content) applicable to *any* physical system[11]. That is, the information content of a physical system can never, he claims, exceed one quarter of the area of its encompassing surface. The black hole saturates the Bekenstein bound, and represents the maximum amount of information that can be packed into the volume occupied by the hole, as befits the equilibrium end state of a gravitating system. A simple argument in support of the universal Bekenstein bound is that if a system confined to a certain region of space possessed an information content in excess of the bound, one could then add some matter and induce this system to undergo gravitational collapse to a black hole, thereby reducing its entropy and violating the second law of thermodynamics (suitably generalized to include event horizon area). However, the Bekenstein bound remains a conjecture: a general proof is lacking.

The idea of associating entropy and information with horizon area was soon extended to include *all* event horizons, not just those surrounding black holes. For example, if the universe becomes dominated by dark energy, which is what current astronomical observations suggest, it will continue to expand at an accelerating rate (dark energy acts as a sort of antigravity force). This creates a *cosmological* event horizon, which may be envisaged as a roughly spherical surface that bounds the region of the universe to which we can ever have causal and informational access. A similar horizon characterizes the period of inflation, widely believed to have occurred at about $10^{-34}$ s after the big bang. Generalizations of horizon entropy have been proposed for cosmological horizon area too, with de Sitter space (a universe subject to dark energy alone) saturating the Bekenstein bound, by Gibbons and Hawking[12], Bousso[13], and Davis and Davies[14]. A number of calculations support the proposal.

Based on the foregoing ideas, 't Hooft[15] and Susskind[16] have proposed the so-called *holographic principle*, according to which the information content of the entire universe is captured by an enveloping surface that surrounds it. The principle states that the total information content of a region of space cannot exceed one quarter of the surface area that confines it (other variants of the holographic principle have been proposed, with different definitions of the enveloping area), and that this limit is attained in the case of the cosmological event horizon. A simple calculation of the size of our universe's event horizon today based on the size of the event horizon created by the measured value of dark energy gives an information bound of $10^{122}$ bits, the same as found by Lloyd using the particle horizon. The event horizon also expands with time, and at this epoch is roughly the same radius as the particle horizon, but unlike the latter, it asymptotes to a constant value not a lot greater than its present value (assuming that the density of dark energy is constant). So whether we take the particle horizon or the event horizon, or a more generalized holographic principle, as the basis for the calculation, we discover an upper bound like (1) on the information content of a causal region of the universe.

How might the bound affect physics and cosmology? The answer to this question depends critically on one's assumptions about the nature of information. The traditional logical dependence of laws, states of matter and information is

$$\text{A. laws of physics} \rightarrow \text{matter} \rightarrow \text{information.}$$

Thus, conventionally, the laws of physics form the absolute and eternal bedrock of physical reality and, as mentioned, cannot be changed by anything that happens in the universe. Matter conforms to the "given" laws, while information is a derived, or secondary property having to do with certain special states of matter. But several physicists have suggested that the logical dependence should really be as follows:

$$\text{B. laws of physics} \rightarrow \text{information} \rightarrow \text{matter.}$$

In this scheme, often described informally by the dictum "the universe is a computer," information is placed at a more fundamental level than matter. Nature is regarded as a vast information-processing system, and particles of matter are treated as special states which, when interrogated by, say, a particle detector, extract or process the underlying quantum state information so as to yield particle-like results. It is an inversion famously encapsulated by Wheeler's pithy phrase 'It from bit'[17]. Treating the universe as a computer has been advocated by Fredkin[18], Lloyd[8] and Wolfram[19] among others.

An even more radical transformation is to place *information* at the base of the logical sequence, thus

$$\text{C. information} \rightarrow \text{laws of physics} \rightarrow \text{matter.}$$

The attraction of scheme C is that, after all, the laws of physics are informational statements.

For most purposes the order of logical dependence does not matter much, but when it comes to the information bound on the universe, one is forced to confront the status of information: is it ontological or epistemological? If information is simply a description of *what we know* about the physical world, as is implied by Scheme A, there is no reason why Mother Nature should care about the limit (1). Or, to switch metaphors, the bedrock of physical reality according to Scheme A is sought in the perfect laws of physics, which live elsewhere, in the realm of the gods – the Platonic domain they are held by tradition to inhabit – where Mother Nature can compute to arbitrary precision with the unlimited quantity of information at her disposal. According to orthodoxy, the Platonic realm is the "real reality," while the world of information is but the shadow on Plato's cave. But if *information* underpins physical reality – if, so to speak, it occupies the ontological basement – (as is implied in Scheme C and perhaps B) then the bound on $I_{\text{universe}}$ represents a fundamental limitation on *all reality*, not merely on states of the world that humans perceive.

Someone who advocated precisely this latter position was Rolf Landauer, a former colleague of Chaitin's at IBM. He explicitly took the view that "the universe computes in the universe," because he believed, as he was fond of declaring, that "information is physical." And Landauer was quick to spot the momentous consequences of this shift in perspective:

"The calculative process, just like the measurement process, is subject to some limitations. A sensible theory of physics must respect these limitations, and should not invoke calculative routines that in fact cannot be carried out."[20]

In other words, in a universe limited in resources and time – a universe subject to the information bound (1) in fact – concepts like real numbers, infinitely precise parameter values, differentiable functions, the unitary evolution of a wave function – are a fiction: a useful fiction to be sure, but a fiction nevertheless, and with the potential to mislead. It then follows that the laws of physics, cast as idealized infinitely precise mathematical relationships inhabiting a Platonic heaven, are also a fiction when it comes to applications to the real universe. Landauer's proposal that our theories should be constrained by the – possibly finite – resources of the universe has been independently developed in recent years by Benioff[21].

If one adopts Landauer's philosophy, then some serious consequences follow. In effect, one cannot justify the application of the laws of physics in situations employing calculations that involve numbers greater than about $10^{122}$, and if one does, then one might expect to encounter departures between theory and experiment. What might this mean in practice? Well, for many purposes, the information bound is so large that the consequences are negligible. Take for example the law of conservation of electric charge. If this were to fail at the $10^{122}$ bit level of accuracy, the implications are hardly dire. (The law has been tested only to about one part in $10^{12}$.)

There are situations in which very large numbers routinely crop up in theoretical physics calculations. One obvious class of cases is where exponentiation occurs. Consider, for example, statistical mechanics, where Poincaré recurrence times are predicted to be of order $\exp(10^N)$ Planck times (chosen to make the number dimensionless) and $N$ is the number of particles in the system. Imposing a bound of $10^{122}$ implies that the recurrence time prediction is reliable only for recurrence times of about $10^{60}$ years. Again, this is so long we would be unlikely to notice any departure between theory and observation. Closely related is the problem of Laplace's demon already discussed. Imposing the information bound renders the demonic prediction valueless almost immediately, because the bound will be exhausted after of order one bit-flip of the $10^{122}$ degrees of freedom in the universe. Exponentiation arises in chaos theory too, via the Lyapunov coefficient. In these examples, the fact that the underlying deterministic mechanics might possess only finite precision is of little importance, because any uncertainties thereby generated are already totally swamped by the practical breakdown of predictability involved in complex and/or chaotic systems.

A case of exponentiation in a relatively simple system occurs in general relativity in connection with the formation of event horizons. For example, when a star implodes to form a black hole, light leaving the surface of the star is exponentially redshifted with an e folding time typically of order a few microseconds. What happens, then, if the exponential redshift is cut off at $10^{122}$ Planck lengths? The classical properties of the black hole are scarcely affected, but Hawking's original derivation of black hole radiance is invalidated, as it already well known[22].

Inflation in the very early universe involves an exponential rate of expansion, i.e. a de Sitter phase, and this offers a stringent test of the information bound hypothesis. It is a key feature of the information bound that it is time-dependent. In the past, the bound was smaller, and its effects on physics would have been greater (see Eq.(2)). During the very early universe, the effects could have been significant, and may have left a trace on the structure of the universe that could be used to test the existence of the bound. Inflation is a brief episode of exponential expansion thought to have occurred at about $10^{-34}$ s after the big bang. At that time, the horizon size was about $3\times10^{-24}$ cm, yielding a surface of about $10^{-19}$ Planck areas. The information bound then implies for the cosmological scale factor change

$$a(t_{after})/a(t_{before}) < 10^{19}. \tag{5}$$

Guth's original proposal was for an inflation factor at least $10^{20}$, so (given the rough-and-ready nature of the calculation) the information bound is consistent with inflation, but only marginally so, and a more detailed analysis may suggest observable consequences, such as a measurable departure from spatial flatness.

Another class of problems in which large numbers are unavoidable is quantum mechanics and quantum field theory, and it is to that topic that I now turn.

3. The quantum vacuum

In quantum mechanics the state of the system is described by a vector in a Hilbert space. For a generic problem, the Hilbert space will possess an infinite number of dimensions. Clearly this construction comes into conflict with the information bound hypothesis. A simple example of the problem concerns the energy of the quantum vacuum, evaluated by summing zero point modes over an infinite set of simple harmonic oscillators[23]. For a massless scalar field confined to a cube of space of linear dimension $L$, the energy density $\rho$ of the vacuum is given by

$$\rho = \tfrac{1}{2}\hbar c L^{-1} \sum_{\mathbf{k}} \omega, \tag{5}$$

where the sum is taken over all the field modes of momentum k. The right hand side of Eq. (5) diverges like $\sim \omega^4$ as $\omega \to \infty$. It may be rendered finite by imposing a cut-off in the summation. A natural cut-off is provided by the Planck frequency, which incorporates only the fundamental constants already present in the theory: $\hbar$, $c$ and $G$. Using this cut-off, Eq. (5) yields a vacuum energy density of $10^{113}$ Jm$^{-3}$, which is some $10^{122}$ times the observed dark energy density. This staggering discrepancy between theory and observation has been known for many years, and is known as the dark energy (or cosmological constant) problem. It is one of the main outstanding challenges to physical theory.

The occurrence of the same factor $10^{122}$ in this discrepancy as in the cosmological information bound is a clear pointer to an alternative explanation for dark energy, and indeed, inequality (1) provides a second natural cut-off for the summation in Eq. (5). Rewriting (5) in terms of modes,

$$\rho \approx \hbar c L^{-4} \sum n^4. \tag{6}$$

If it is now argued that the sum $\sum n^4$ should be bounded by (1), then taking $L$ to be the horizon radius (roughly a Hubble radius) and $\sum n^4 \sim 10^{122}$, we may evaluate the vacuum energy density to be

$$\rho \approx 10^{-9} \text{ Jm}^{-3} \approx \rho_{\text{observed}}. \tag{7}$$

The same result may be derived in a completely different way, by imposing the condition on the vacuum energy that at every scale of size $L$, the energy density must not exceed the level at which the total mass within a volume $L^3$ is greater than the mass of a black hole of size $L$, otherwise the vacuum energy would presumably undergo gravitational collapse. This requirement may be expressed as follows:

$$\rho c^2 L^3 < M_{\text{bh}}(L). \tag{8}$$

Substituting the right hand side of Eq. (5) for $\rho$ we obtain, to an order or magnitude,

$$G\hbar\omega^4 L^3/c^7 < L \tag{9}$$

or

$$\rho < c^4/GL^2 \tag{10}$$

Taking $L$ to be the Hubble radius, inequality (10) may be re-cast in the following suggestive form[24]:

$$\rho < (\rho_P \rho_H)^{\frac{1}{2}} \approx 10^{-9} \text{ Jm}^{-3} \approx \rho_{\text{observed}} \tag{11}$$

where $\rho_P$ is the Planck energy density and $\rho_H$ is the Hubble energy density, defined to be the energy density of a single quantum in a Hubble volume with a wavelength equal to the Hubble radius.

This remarkable result – that the cosmological information bound explains the magnitude of the dark energy – comes at a price, however. The same reasoning may be applied to the *pressure* of the vacuum, $p$, which for a massless scalar field is

$$p = -\tfrac{1}{2}\hbar c L^{-1} \sum \omega, \tag{12}$$

i.e. $p = -\rho$, which is the necessary equation of state for the vacuum energy to play the role of dark energy. Now recall that the information bound varies with time in the manner indicated by Eq. (2). Hence the cut-off in the summation in both Eqs. (5) and (12) will be time-dependent, so the dark energy is also predicted to be time-dependent. This raises an immediate difficulty with the law of energy conservation:

$$p\, da^3 + d(\rho a^3) = 0 \tag{15}$$

which can be satisfied for a time-dependent $p$ and $\rho$ only if there is some compensatory change, e.g. $G$ and/or $c$ vary with time. There is a substantial literature on such holographic cosmological models[25], including comparison with observations, which I shall not review here.

## 4. Quantum information processing

As a final application of the information bound hypothesis, let me turn to non-relativistic quantum mechanics. A transformation in our understanding of information came with the recognition that because nature is fundamentally quantum mechanical, the rules for information processing at the quantum level differ not only in the technical details but in their very conceptual basis from the classical case. In conventional (classical) information theory, the basic unit is the bit, or binary choice, usually symbolized by 0 and 1. In quantum mechanics, the bit is replaced by a more abstract entity: the qubit. When humans read out the information content of a quantum system, they appropriate only bits – the act of read-out collapses qubits into bits. But the importance of quantum information dynamics is that in an isolated unobserved quantum system, the qubits generally evolve in a manner completely different from the classical case, involving the whole panoply of quantum weirdness, including, most crucially, superposition and entanglement. It is this feature that has commended quantum information science to governments and business by holding out the promise of large-scale quantum computation. By exploiting qubit dynamics, a quantum computer would represent an unprecedented leap in computational power[26].

The key to quantum computation lies with the exponential character of quantum states. Whereas a classical binary switch is either on (1) or off (0), a quantum system can be in a superposition of the two. Furthermore, a multi-component quantum system can incorporate entanglement of spatially separated subsystems. Combining these two properties implies that an $n$-component system (e.g. $n$ atoms) can have $2^n$ states, or components of the wave function, that describe the system. If it were possible to control all the components, or branches, of the wave function simultaneously, then the quantum system would be able to process information exponentially more powerfully than a classical computer. This is the aspiration of the quantum computation project.

Because the complexity of an entangled state rises exponentially with the number of qubits (which is its virtue), large-scale quantum information processing comes into conflict with the information bound. Specifically, a quantum state with more components than about $n = \log_2 I_{universe}$ will require more bits of information to specify it than can be accommodated in the entire observable universe! Using the bound given by inequality (1), this yields a limit of approximately $n = 400$. In other words, a generic entangled state of more than about 400 particles will have a quantum state with more components than $I_{universe}$, evolving in a Hilbert space with more dimensions than $I_{universe}$. The question therefore arises of whether this violation of the information bound (1) signals a fundamental physical limit. It seems to me that it must.

On the face of it, the limit of 400 particles is stringent enough to challenge the quantum computation industry, in which a long-term objective is to entangle many thousands or even millions of particles and control the evolution of the quantum state to high precision. The foregoing analysis, however, is overly simplistic. First, note that the dimensionality of the (non-redundant part of the) Hilbert space is not an invariant number: by changing the basis, the number might be reduced. So specifying the complexity of a quantum state simply by using the dimensionality of the Hilbert space can be misleading. A more relevant criterion is the number of independent parameters needed to specify inequivalent $n$-component quantum systems. This problem has been addressed, but it is a difficult one on which only limited progress has so far been made[27].

Second, the dimensionality of the Hilbert space serves to define the number of amplitudes needed to specify a generic superposition. But the amplitudes themselves require additional information to specify them; indeed, a single complex number coefficient $α_i$ will mostly contain an infinite number of bits of information. If we are to take the bound (1) seriously, then it must be applied to the total algorithmic information content of the amplitude set over the entire Hilbert space. Following Chaitin, the algorithmic information measure of a binary string $X$ is defined as

$$H(X) = - \ln P(X) + O(1) \qquad (16)$$

where $P(X)$ is the probability that the proverbial monkey typing randomly on a typewriter will generate a program which, when run on a universal Turing machine, will output $X$. Applied to the amplitude set $\{α_i\}$ of a generic quantum state (plus any ancillary information needed to specify the state, such as constraints), the cosmological information bound (1) may be expressed as follows:

$$H(\{α_i\}) < A_{holo}/L_P^2 \qquad (17)$$

where $A_{holo}$ is the area of the appropriate holographic surface (e.g. a cosmological event horizon). Inequality (17) is a stronger constraint than (1), appropriate to the interpretation of information as ontological and fundamental, and therefore including not merely a head-count of the degrees of freedom, but the algorithmic information content of all the specifying parameters of the state too. This extra informational burden on the bound will reduce somewhat the dimensionality of the Hilbert space at which unitary evolution is expected to break down.

A more subtle issue concerns the specific objectives of quantum computation, which is not to control the dynamical evolution of *arbitrary* entangled quantum states, but an infinitesimal subset associated with certain mathematical problems of interest, such as factoring. It is trivially true that it is impossible to prepare, even approximately, a state containing more than $10^{122}$ truly independent parameters because it is impossible to even specify such a state: there are not enough bits in the universe to contain the specification. Almost all states fall into this category of being impossible to specify, prepare and control. So in this elementary sense, generic quantum computation is obviously impossible. Less obvious, however, is whether the subset of states (of measure zero) of interest to the computing industry is affected by the cosmological information bound, for even if it is the case that the number of independent amplitudes exceeds $10^{122}$, there may exist a compact mathematical algorithm to generate those amplitudes. (The algorithm for generating the amplitudes that specify the initial state should not be confused with the algorithm to be executed by the quantum computer dynamics.) For example, the amplitudes of the quantum computer's initial state could be the (unending) digits of π, which can be generated by a short algorithm. That is, the set of amplitudes may contain an unbounded number of bits of information, but a finite (and even small) number of bits might be sufficient to define the generating algorithm of the amplitude set. So if the information bound on the universe is interpreted as an upper limit on the *algorithmic* information (as opposed to the Shannon information), then a measure-zero subset of initial states can be specified without violating the cosmological information bound. But this loophole leaves many unanswered questions. For example, a mathematical specification is one thing, a physical process to implement that specification – and to do so in an acceptable period of time – is another. To take the cited example, it is far from clear that there exists *any* physical process that can create an entangled quantum state in which the amplitudes (enumerated in some sequence) are the digits of $\pi$. And even if this further problem is satisfactorily addressed, one has to confront the fact that as the initial state evolves, and the amplitudes change, so the set of amplitudes may not remain algorithmically compressible. To be sure, a unitary evolution of an initially algorithmically compressible state will, by definition, preserve algorithmic compressibility (because the unitary operation is an algorithm). But such a pure system is unstable: the inevitability of random errors due to the fact that the quantum system is not closed will raise the algorithmic complexity, and seemingly raise it above the bound (1) in pretty short order[28]. This uncovers a deeper set of issues, which is whether a quantum state that cannot be specified, and is in principle unknowable, and the amplitude set of which exceeds the total information capacity of the universe, may nevertheless still be said to exist and conform to physical law. According to the Landauer point of view I am articulating here, the answer is no.

### 5. Unfinished business

I have been asked what, exactly, would go wrong if one tried to build and operate a quantum computer with, say, 500 entangled qubits. First let me make a general point. In science, one always has to distinguish between mathematical possibility contained in a theory, and physical possibility. For example, general relativity contains mathematical models with closed timelike world lines, but these may be inconsistent with cosmological boundary conditions or some other global requirement[29]. So the fact that a unitary transformation that implements a desirable quantum computation may exist mathematically does not necessarily mean it can be implemented physically, even in principle. And in fact, a prima face example would seem to be the expectation that the resources needed to prepare an initial quantum state are expected to grow with its complexity, and would require more and more of the surrounding universe to be commandeered, and more yet for the error correction of its evolution. Inevitably, the gravitational effects of the commandeered matter will eventually become important. Before the complexity of the state reached the cosmological bound of $10^{122}$, the entire resources of the observable universe would necessarily be exhausted. Thus, almost all quantum initial states, and hence almost all unitary transformations, seem to be ruled out by the cosmological constraint (1) (if one accepts it). It is important to realize, however, that this restriction may not be an impediment to preparing an algorithmically simple state, providing a physical mechanism can be found to implement the preparation algorithm. These criteria will undoubtedly be satisfied for the (very limited) examples of known quantum algorithms, such as Shor's algorithm for factorization, which is algorithmically simple by definition, since its input state can be specified and there is a simple association between the input data and the initial quantum state. What is less clear is whether this ease of preparation of the initial state is representative of a broader class of problems of interest, or remains confined to a handful of special cases.

A more radical conjecture about what might go wrong concerns the subsequent evolution of the state, which entails an escalation of the algorithmic complexity through the cosmological information bound due to random errors in the manner I mentioned above. Under these circumstances, it may be that the unitary evolution of the state actually breaks down (over and above the breakdown caused by tracing out the degrees of freedom associated with the errors caused by environmental disturbances). This would manifest itself in the form of an additional source of errors, ultimately of cosmological origin, in a manner such that all error-correcting protocols applied to these errors would fail to converge. What I am suggesting here seems to be close to the concept of unavoidable intrinsic decoherence proposed by Milburn[30]. Some clarification of these issues may emerge from the further study of the recent discovery that the entropy of quantum entanglement of a harmonic lattice also scales like area rather than volume[31], which would seem to offer support for the application of the holographic principle to entangled states. It would be good to know how general the entanglement-area relationship might be.

Finally, I should point out that the information bound (1) was derived using quantum field theory, but that same bound applies to quantum field theory. Ideally one should derive the bound using a self-consistent treatment. If one adopts the philosophy that information is primary and ontological, then such a self-consistency argument should be incorporated in a larger program directed at unifying mathematics and physics. If, following Landauer, one accepts that mathematics is meaningful only if it is the product of real computational processes (rather than existing independently in a Platonic realm) then there is a self-consistent loop: the laws of physics determine what can be computed, which in turn determines the informational basis of those same laws of physics. Benioff has considered a scheme in which mathematics and the laws of physics co-emerge from a deeper principle of mutual self-consistency[32], thus addressing Wigner's question of why mathematics is so "unreasonably effective" in describing the physical world.[33] I have discussed these deeper matters elsewhere[34].

**Acknowledgments**

I should like to thank Scott Aaronson, Ted Jacobson, Gerard Milburn, William Phillips, Sandu Popescu and Leonard Susskind for helpful comments, conversations and guidance.

**Footnotes**

[1] Wittgenstein, L. (1921) *Tractatus Logico-Philosophicus,* English translation: David Pears and Brian McGuinness (Routledge, London 1961).

[2] Aaronson, S. (2005) 'Are quantum states exponentially long vectors?' *Proceedings of the Oberwolfach Meeting on Complexity Theory* (to be published).

[3] Chaitin, G. (2005) *Meta Math! The Quest for Omega* (Pantheon Books, New York), 115.

[4] *Cosmic Jackpot* by Paul Davies (Houghton Mifflin, New York 2007).

[5] Laplace, P. (1825) *Philosophical Essays on Probabilities* (trans. F.L. Emory and F.W. Truscott, Dover, New York 1985).

[6] I am grateful to Michael Berry for drawing my attention to this example.

[7] 'The limits of reason,' by Gregory Chaitin, *Scientific American* March 2006, p. 74.